\documentclass[12pt]{article}
\usepackage{amsmath,epsfig,graphicx,amssymb}
\usepackage[margin=0.7in]{geometry}
\date{}
\begin{document}
\title{The Quantum-like Face of Classical Mechanics}
\author{Partha Ghose\footnote{partha.ghose@gmail.com} \\
The National Academy of Sciences, India,\\ 5 Lajpatrai Road, Allahabad 211002, India}
\maketitle
\begin{abstract}
It is first shown that when the Schr\"{o}dinger equation for a wave function is written in the polar form, complete information about the system's {\em quantum-ness} is separated out in a single term $Q$, the so called `quantum potential'. An operator method for classical mechanics described by a `classical Schr\"{o}dinger equation' is then presented, and its similarities and differences with quantum mechanics are pointed out. It is shown how this operator method goes beyond standard classical mechanics in predicting coherent superpositions of classical states but no interference patterns, challenging deeply held notions of classical-ness, quantum-ness and macro realism. It is also shown that measurement of a quantum system with a  classical measuring apparatus described by the operator method does not have the measurement problem that is unavoidable when the measuring apparatus is quantum mechanical. The type of decoherence that occurs in such a measurement is contrasted with the conventional decoherence mechanism. The method also provides a more convenient basis to delve deeper into the area of quantum-classical correspondence and information processing than exists at present.
\end{abstract} 

\section{Introduction}
The hallmark of quantum mechanics is the characterization of physical states by vectors in a Hilbert space with observables represented by Hermitean operators that do not all commute. The description is non-realist and non-causal. On the other hand, physical states in standard classical mechanics are points in phase space, all observables commute, and the description is realist and causal. To bridge this gap between quantum and classical mechanics, two different approaches have been made: (i) interpret quantum mechanics as closely as possible to classical mechanics, i.e. give it a realist and causal interpretation \cite{db, bohm}, or (ii) formulate classical mechanics as closely as possible to quantum mechanics \cite{rosen1, rosen2, rosen3, gh1, gh2}. In this paper we will follow the latter approach to explore some of its less known and unexplored aspects. 

Let us start with the Schr\"{o}dinger equation
\begin{eqnarray} 
i\,\hbar\,\frac{\partial \psi(\vec{x},t)}{\partial t} &=& \left[\frac{\hat{p}^2}{2m} + V(\vec{x},t) \right]\psi(\vec{x},t) =\left(-\frac{\hbar^2}{2 m}\, \nabla^2  + V(\vec{x},t)\right)\psi(\vec{x},t) \equiv \hat{H}_{qm}\psi(\vec{x},t) \label{Sch}
\end{eqnarray} 
for a wave function $\psi(\vec{x},t)$ of a system.
Writing $\psi(\vec{x},t)$ in the polar form 
\begin{equation}
\psi(\vec{x},t)= \sqrt{\rho_q(\vec{x},t)}\, {\rm exp}(iS_q(\vec{x},t)/\hbar) \label{polar}
\end{equation}
 and separating the real and imaginary parts, one gets the two equations
\begin{eqnarray} 
\frac{\partial S_q(\vec{x},t)}{\partial t} + \frac{(\nabla S_q(\vec{x},t))^2}{2m} + V(\vec{x},t) + Q &=& 0,\,\,\,\, Q = -\frac{\hbar^2}{2 m}\,\frac{\nabla^2 \sqrt{\rho_q(\vec{x},t)}}{\sqrt{\rho_q(\vec{x},t)}},\label{HJQ} \\
\frac{\partial \rho_q(\vec{x},t)}{\partial t} +
\vec{\nabla}. \left(\,\rho_q(\vec{x},t)\, \frac{\vec{\nabla}S_q(\vec{x},t)}{m}\right) &=& 0 \label{contqm}
\end{eqnarray}
which are {\em coupled} equations. These equations were first derived by Madelung as equations for quantum hydrodynamics \cite{madelung}. We will not adopt that interpretation and view them just as mathematical transformations of the Schr\"{o}dinger equation from the complex function $\psi$ to the two real variables $\sqrt{\rho_q}$ and $S_q$. 

Now, consider the classical Hamilton-Jacobi equation
\begin{eqnarray} 
\frac{\partial S(\vec{x},\vec{p}_0,t)}{\partial t} + \frac{(\nabla S(\vec{x},\vec{p}_0,t))^2}{2m} + V(\vec{x},t) &=& \frac{\partial S(\vec{x},\vec{p}_0,t)}{\partial t} + H = 0\label{clHJ}
\end{eqnarray}
for the action $S(\vec{x},\vec{p}_0,t)$ where $\vec{p}_0$ is the initial value of the momentum,  and the Liouville equation
\begin{eqnarray} 
\frac{\partial \rho(\vec{x},\vec{p},t)}{\partial t} + \sum_i \left(\dot{x}_i\frac{\partial}{\partial x_i} + \dot{p}_i\frac{\partial}{\partial p_i}\right)\rho(\vec{x},\vec{p},t) = 0\label{Liouv}
\end{eqnarray} 
for the density $\rho(\vec{x},\vec{p},t)$ in phase space. 
Comparing these two classical equations with the Madelung equations, it becomes clear that, like the quantum mechanical action $S_q$, the classical action $S$ is also a function of only $(\vec{x},t)$, $\vec{p}_0$ being a parameter fixed by the initial momentum. However, unlike the quantum mechanical probability density $\rho_q$ which is a function of only $(\vec{x},t)$, the  classical phase space density $\rho$ is a function of $(\vec{x},\vec{p},t)$. Hence, before a meaningful comparison between these two sets of equations can be made, one must take the projection of $\rho(\vec{x},\vec{p},t)$ to the coordinate subspace density $\rho(\vec{x},t)$.
It follows from the classical equations (\ref{clHJ}) and (\ref{Liouv}) that these functions satisfy the equations
\begin{eqnarray} 
\frac{\partial S(\vec{x},t)}{\partial t} + \frac{(\nabla S(\vec{x},t))^2}{2m} + V(\vec{x},t) &=& 0,
\label{1}\\
\frac{\partial \rho(x,t)}{\partial t} + \vec{\nabla}.(\rho(x,t)\vec{v}) = 0.\label{2}
\end{eqnarray}
Now comparing this set with the Madelung set, one can clearly see that the first equation of the Madelung set differs from the first equation of the classical set by the presence of a single term $Q$. It is this term therefore which encapsulates in it all quantum mechanical features, and also couples the Madelung set: $S_q$ gets a $\sqrt{\rho_q}$ dependence through $Q$. However, unlike the two Madelung equations, the two classical equations are {\em not} coupled. 
  
The question that naturally arises is: is there an equation whose real and imaginary parts are the two classical equations (\ref{1}) and (\ref{2})? There indeed is such an equation, and it is \cite{rosen3, gh1, gh2}
\begin{eqnarray}  
i\,\hbar\,\frac{\partial \psi_{cl}(\vec{x},t)}{\partial t} &=& \left[\hat{H}_{qm} - Q_{cl}\right]\psi_{cl}(\vec{x},t) \equiv \hat{H}_{cl}\psi_{cl}(\vec{x},t), \label{Sch2}\\
Q_{cl} &=& -\frac{\hbar^2}{2 m}\,\frac{\nabla^2 \sqrt{\rho(\vec{x},t)}}{\sqrt{\rho(\vec{x},t)}}
\end{eqnarray} 
for a complex classical wave function 
\begin{equation}
\psi_{cl}(\vec{x},t) = \sqrt{\rho(\vec{x},t)}\,{\rm exp}(iS(\vec{x},t)/\hbar).\label{clwave}
\end{equation}
It is known as the ``classical Schr\"{o}dinger equation''. Although this equation involves $\hbar$, it drops out of the classical equations (\ref{1}, \ref{2}). Hence, put in this Schr\"{o}dinger-like form, classical mechanics does not require $\hbar$ to be small compared to $S$ or to vanish.

If one imposes the normalization condition
\begin{equation}
\int|\psi_{cl}(\vec{x},t)|^2 d^3 x = \int \rho(\vec{x},t)\, d^3 x = 1, 
\end {equation} 
$\psi_{cl}(\vec{x},t)$ can be interpreted as a position probability amplitude in configuration space, just like the quantum wave function $\psi_q(\vec{x},t)$. 

Formulated in this way, classical states are no longer points in phase space but vectors in a Hilbert space spanned by complex square integrable classical wave functions and acted on by non-commuting operators like $\hat{p} = -i\hbar \nabla$ and $x$. This opens up new possibilities in classical mechanics that are worth exploring.

\section{Similarities and Differences between Quantum and Classical Systems}

(i) The classical Schr\"{o}dinger equation is remarkably similar to the Schr\"{o}dinger equation. The only difference between them lies in the difference between the two Hamiltonians $\hat{H}_{qm}$ and $\hat{H}_{cl}$: 
\begin{eqnarray}
\hat{H}_{cl} &=& \hat{H}_{qm} - Q_{cl} = \hat{H}_0 + V - Q_{cl},\\
\hat{H}_0 &=& \frac{\hat{p}^2}{2m}.
\end{eqnarray}
If the external potential $V=0$, the free classical Hamiltonian operator is
\begin{equation}
\hat{H}_{cl\,0} = \hat{H}_0 - Q_{cl}.
\end{equation}
Notice that
\begin{equation}
Q_{cl} \sqrt{\rho} = -\frac{\hbar^2\nabla^2 \sqrt{\rho}}{2 m}  = \frac{\hat{p}^2}{2m}\sqrt{\rho} = \hat{H}_0 \sqrt{\rho},\,\,\,\, \hat{p} = -i\hbar \nabla, \label{phat}
\end{equation}
showing that $Q_{cl}$ is, in fact, the kinetic energy operator acting on the amplitude $\sqrt{\rho}=|\psi_{cl}|$ of the wave function, and is hence Hermitean when operating on $|\psi_{cl}|$. Writing
\begin{eqnarray}
\hat{H}_0\psi_{cl} &=& (\hat{H}_0 \sqrt{\rho})\,e^{iS/\hbar} + \sqrt{\rho} (\hat{H}_0 e^{iS/\hbar}),\nonumber\\
Q_{cl}\psi_{cl} &=& (\hat{H}_0 \sqrt{\rho})\,e^{iS/\hbar}, 
\end{eqnarray}
one gets\begin{eqnarray}
\hat{H}_{cl\,0}\psi_{cl} &=& (\hat{H}_0 - Q_{cl})\psi_{cl} = \sqrt{\rho} \hat{H}_0 e^{iS/\hbar}.
\end{eqnarray}
This shows that the nonlinear term $Q_{cl}$ exactly cancels the quantum mechanical kinetic energy of the amplitude $\sqrt{\rho}$, and the free classical Hamiltonian operator acts only on the phase of the classical wave function.

Hence, to quantize a classical system, one has to put the term $Q_{cl} = 0$ in the classical Hamiltonian operator $\hat{H}_{cl}$. On the other hand, to `classicalize' a quantum system, one has to subtract the term $Q_{cl}$ from the quantum Hamiltonian $\hat{H}_{qm}$. Thus, the quantity $Q_{cl}$ plays the crucial role of coupling or decoupling the amplitude and the phase of a wave function and thus create the difference between quantum and classical wave functions.
\vskip0.1in
(ii) Being a functional of $|\psi_{cl}|$, $Q_{cl}$ is a {\em nonlinear} operator $Q_{cl}\hat{\mathbb{I}}$ when operating on $\psi_{cl}$. Hence, the classical Schr\"{o}dinger equation is nonlinear, and superpositions of different solutions of the equation are not solutions of the equation in general. However, there is one exception. If $\psi_{cl}$ is a solution, $\alpha \psi_{cl}$ is also a solution provided $\alpha$ is a real or complex function independent of $|\psi_{cl}|$. Now, note that one can write $\alpha = 1 + e^{i\theta(x,t)}$. Hence, the following Lemma holds:
\vskip0.1in
{\flushleft{{\em Lemma}}}
\vskip0.1in
Superpositions of solutions with the same amplitude $\sqrt{\rho}$ but different phases are also solutions of the classical Schr\"{o}dinger equation.
\vskip0.1in
(iii) Since $[x, \hat{p}]= i\hbar$, the standard deviations $\sigma_x = \sqrt{\langle \hat{x}^2\rangle - \langle \hat{x}\rangle^2}$ and $\sigma_p = \sqrt{\langle \hat{p}^2\rangle - \langle \hat{p}\rangle^2}$ with the expectation values taken in classical wave packets satisfy the Robertson uncertainty relation $\sigma_x \sigma_y \geq \hbar/2$ \cite{rob}. This characterizes {\em inherent fluctuations} of the observables in ensembles and does not reflect trade-off relations between errors and disturbances in measurement processes related to Heisenberg's uncertainty principle \cite{busch, ozawa} which does not operate in classical systems. 
\vskip0.1in 
(iv) For time independent classical Hamiltonians one can write
\begin{equation}
S(\vec{x}, t) = W(\vec{x}) - Et = \vec{p}.\vec{x} - Et\label{action2} 
\end{equation}
such that $\vec{\nabla}S(\vec{x}, t) = \vec{\nabla}W(\vec{x}) = \vec{p}$.
The phase factor ${\rm exp}(iS(\vec{x},t)/\hbar)$ of a classical wave function can be written in the standard form ${\rm exp}[i(\vec{k}.\vec{x} -\omega t)]$ provided $\vec{p} = \hbar \vec{k},\,E = \hbar \omega$. Hence, the dispersion relation is $\omega = \hbar k^2/2m$, and the phase velocity is $|v|_\varphi = \omega/k = \hbar k/2m$ which is different for different wavelengths $\lambda = h/p = 2\pi/k$, exactly as in quantum mechanics. It follows therefore that classical wave packets for material particles also spread in general. 
\vskip0.1in
(v) The functions $S(\vec{x},t)$ and $\sqrt{{\rho}(\vec{x},t)}$ are defined for a fixed momentum $\vec{p}_0$. Furthermore, they evolve independent of each other. Consequently, trajectories or `rays' generated by solutions $x(t)$ of the first-order differential equation 
\begin{equation}
d\vec{x}/dt = \frac{1}{m}\vec{\nabla} S(\vec{x},t)\label{traj1}
\end{equation}
contain an arbitrary constant of integration determined by the initial positions, but they all have the same momentum $\vec{p}_0$. They do not cross because of the single-valuedness of the wave function \cite{ben}. This equation, however, does not have information about how $S$ evolves dynamically. That is contained in eqn (\ref{1}) from which Newton's second law of motion 
\begin{equation}
\frac{d\vec{p}}{dt} = - \vec{\nabla}V\label{newton}
\end{equation}
can be derived by taking its gradient and putting $\vec{\nabla} S(\vec{x}, t) = \vec{p}$. This is a first-order differential equation in $p$, and its solutions $p(t)$ contain an arbitrary constant determined by the initial momenta. They, however, all have the same position. These two first-order equations are basically the pair of Hamilton equations
\begin{eqnarray}
\dot{x}_i &=& \frac{\partial H}{\partial p_i},\,\,\,\,p_i = \nabla_i S,\\
\dot{p}_i &=& -\frac{\partial H}{\partial x_i},
\end{eqnarray}
which together determine the classical trajectories, given both the initial positions and momenta. The complete dynamical description of such trajectories is therefore given by the phase space distribution $\rho(\vec{x},\vec{p},t)$ and the Liouville equation, and not by the first-order equation (\ref{traj1}) alone, as assumed in Refs \cite{rosen3, ben}.
\vskip0.1in  
(vi) In the causal interpretation of quantum mechanics \cite{bohm}, too, the trajectories determined by the first-order equation $\vec{p} = m d\vec{x}/dt = \vec{\nabla} S_q(\vec{x},t)$ (the guidance condition) do not cross because of the single-valuedness of the wave function, and are `hidden variables'. In this interpretation Newton's second law gets modified to the form
\begin{equation}
\frac{d\vec{p}}{dt} = m \frac{d^2\vec{x}}{dt^2} = - \vec{\nabla}(V + Q),
\end{equation}
and the trajectories determined by solutions of this second-order equation in $x$ with the initial position distribution fixed to match the quantum mechanical distribution $|\psi(\vec{x},t)|^2_{t=0}$ {\em and} restricting the initial momenta to satisfy the guidance condition, also do not cross and are `hidden'. It is clear that the term $Q$ accounts for the differences with classical dynamics.
\vskip0.1in  
(vii) Since the two classical equations (\ref{1}, \ref{2}) are independent of each other, {\em the phase $S/\hbar$ and the amplitude $\sqrt{\rho}$ of a classical wave function $\psi_{cl} = \sqrt{\rho}\,{\rm exp}\,(iS/\hbar)$ evolve independent of each other}. This guarantees the absence of interference patterns in classical systems in spite of the fact that they can be described by a complex wave function. 
\vskip0.1in
An example will now be given to show precisely how the coherent superposition of two classical wave functions does not result in an interference pattern.

\section{Coherence in Classical Systems}
Consider now a typical double-slit set up in the $xy$ plane with two localized classical wave packets 
\begin{eqnarray}
\psi_{cl}(x=0,y=y_1,t=0) &=& \int_{-\infty}^{+\infty}\sqrt{\rho(k_y)} e^{i k_y y_1}dk_y,\\
\psi_{cl}(x=0,y=y_2,t=0) &=& \int_{-\infty}^{+\infty}\sqrt{\rho(k_y)}e^{i k_y y_2}dk_y,
\end{eqnarray}
centred on the two narrow slits at ($y_1, y_2$) at $t=0$. In accordance with the Lemma stated in (ii) above, the wave function at a point $y$ on a detector screen placed at a far enough distance $x$ from the slits where the two wave packets overlap can be written as (ignoring an unphysical overall phase) 
\begin{eqnarray}
\psi_{cl}(x,t) &=& \frac{1}{\sqrt{2}}\int_{-\infty}^{+\infty}\sqrt{\rho(k_x,k_y)}e^{i(k_x x-\omega t)}\left[1 + e^{i\varphi_{cl}}\right]dk_xdk_y\label{clsup}\\
\varphi_{cl} &=&  \int_{y_1}^y k_y dy -  \int_{y_2}^y k_y dy\nonumber\\
&=& \int_{y_1}^{y_2} k_y dy.
\end{eqnarray}
The relative phase $\varphi_{cl}$ is clearly independent of $y$, the position on the detector screen. This is because $\varphi_{cl}$ evolves according to the classical Hamilton-Jacobi equation (\ref{1}) which is independent of the amplitude $\sqrt{\rho}$ which contains all `path' or flow information determined by the continuity eqn (\ref{2}). That is why no interference fringe pattern is formed on the screen. In quantum mechanics (as well as in classical optics), this is not the case, the relative phase at every point $y$ being dependent on the difference between the `paths' from the two slits to the point. This is due to the fact that the Madelung equations are coupled, and the phase is path dependent. However, $\varphi_{cl}$ can be varied by varying the distance $d = (y_2 - y_1)$ between the slits, and a variation of the probability distribution of particles with $d$ on the detector screen is predicted. 

This provides a new insight into the difference between quantum and classical coherence. The former gives rise to interference fringes but not the latter.

Leggett and Garg \cite{lgi} established a contradiction between quantum mechanics and the twin principles of {\em macroscopic realism} and {\em non-invasive measurability on the macroscopic scale}. The former requires that a system with two or more distinct states accessible to it will at all times {\em be} in one or the other of these states. Violations of Leggett-Garg inequalities are supposed to signal quantum behaviour. The state (\ref{clsup}) violates the principle of macroscopic realism, and yet it is non-quantum mechanical. It, however, satisfies the principle non-invasive measurability on the macroscopic scale because there is no Heisenberg uncertainty principle for systems described by the classical Schr\"{o}dinger equation (point (iii) above). Although no such states are known to exist, they are an inevitable consequence of the classical Schr\"{o}dinger equation which is a complex combination of two well known classical equations. Hence, the non-occurrence of such states is puzzling and probably indicates their extreme instability. It must, however, be noted that coherent superpositions of classical states are different from macroscopically coherent quantum mechanical states such as those in SQUIDs because the relative phases of the former are amplitude independent constants.

\section{Measurement} 
Finally, let us consider the case of an observation designed to measure some observable $\hat{P}$ of a quantum particle with wave function $\psi(\vec{x},t)$. Let the wave function of the apparatus be $\psi_{cl}(y,t)$ where $y$ is the relevant coordinate of the apparatus. The initial state is a product state
\begin{equation}
\Psi(\vec{x},y,0) = \psi(\vec{x},0)\psi_{cl}(y,0) = \psi_{cl}(y,0)\sum_p c_p \psi_p(\vec{x},0)
\end{equation}
where $\hat{P}\psi_p(\vec{x}) = p\psi_p(\vec{x})$.

The measurement interaction Hamiltonian is taken to be of the von Neumann type \cite{vn}, namely
\begin{equation}
\hat{H}_I = -g\hat{P}\hat{p}_y
\end{equation}
where $g$ is a suitable constant and $\hat{p}_y$ is the momentum operator corresponding to the coordinate $y$ of the apparatus. Following von Neumann, we assume that during this impulsive interaction $0 < t < \tau$ the free evolution of the particle and the apparatus can be ignored (i.e. $\hat{H}_0 \simeq 0,\, \hat{H}_{cl,0} \simeq 0$ because the masses $m$ and $M$ of the particle and the apparatus respectively are very large), and
\begin{eqnarray}  
i\hbar\frac{\partial \Psi}{\partial t} &=& \hat{H}_I \Psi = -g\hat{P}\hat{p}_y\Psi = ig\hbar\hat{P}\frac{\partial \Psi}{\partial y} \label{mi}
\end{eqnarray}  
is a very good approximation. Thus, {\em during the measurement interaction the total system evolves effectively quantum mechanically with $\hat{H}_I$ as the Hamiltonian} \cite{note}. Let the state at a time $t < \tau$ be  
\begin{equation}
\Psi(x,y,t) = \sum_p c_p\psi_p(\vec{x}) f_p(y,t)
\end{equation}
where $f_p(y,t) = \sqrt{\rho_p}\,{\rm exp}(iS_p(y,t)/\hbar)$ are determined by eqn (\ref{mi}) to satisfy the condition
\begin{equation}
\frac{\partial f_p(y,t)}{\partial t} = gp\frac{\partial f_p(y,t)}{\partial y}
\end{equation}
for each eigenvalue $p$. 
If the initial value of $f_p(y,t)$ is $f_p^0(y)$, then it follows from this that
\begin{equation}
f_p(y,t) = f^0_p (y - gpt).
\end{equation}
Hence, the final wave function is of the entangled form
\begin{equation}
\Psi(\vec{x}, y,t) = \sum_p c_p \psi_p(\vec{x})f^0_p (y - gpt).\label{ent}
\end{equation}
This superposition of the quantum and classical wave functions is valid only for sufficiently small values of $t < \tau$ during which the wave packets $f_p$ correlated to different eigenvalues $p$ of the quantum particle evolve according to eqn (\ref{mi}), i.e. effectively quantum mechanically, and overlap considerably. If one assumes that these wave packets are Gaussians,
\begin{equation}
|f_p(y- gpt)| = \frac{1}{\sigma \sqrt{2\pi}}{\rm exp}\left[- \frac{(y - gpt -y_0)^2}{2\sigma^2}\right],
\end{equation} 
one can use the criterion that two adjacent wave packets are just resolved if their maxima are separated by $2\sqrt{2 ln 2} \approx 2.355 \sigma$, the Full Width At Half Maximum (FWHM). This means that eqn (\ref{ent}) holds for $t < \tau = 2.355\sigma/g\delta p$ where $\delta p$ is the difference between two successive eigenvalues. During this time interval, usually extremely short for sharp Gaussians and large values of $g$, the $f_p$s are correlated to the different eigenvalues $p$ of the quantum mechanical particle, and $\Psi(\vec{x}, y,t)$ is an entangled state. However, for $\tau \geq 2.355\sigma/g\delta p$ these wave packets are resolved and evolve according to the classical Schr\"{o}dinger equation, and consequently their sum is no longer a solution of the classical Schr\"{o}dinger equation which is nonlinear. Hence, finally there is no single wave function $\Psi(x,y,t)$ for the state, which evolves to a mixed state with the Born probabilities $|c_p|^2$ for each eigenvalue $p$. The transition from the pure to the mixed state takes place dynamically through the classical Schr\"{o}dinger equation, and the time taken is shorter the stronger the coupling $g$ \cite{note}. There is therefore no need for a projection/collapse postulate, and hence there is no measurement problem in the theory.

For Bell states like $\frac{1}{\sqrt{2}}[|0\rangle_A|0\rangle_B \pm |1\rangle_A|1\rangle_B]$ the classical measuring apparatus states $|\psi\rangle_{cl\,A}$ and $|\psi\rangle_{cl\,B}$ on the two sides evolve to mixed states with definite probabilities for the two possible outcomes correlated to the states $|0\rangle_A,|1\rangle_A$ on one side and to the states $|1\rangle_B,|0\rangle_B$ on the other side. The observations reveal nonlocal correlations but there is no collapse and no action-at-a-distance.
\subsection{Measurement and Decoherence}
The process of decoherence involved in measurements with a classical apparatus is of a different type from conventional decoherence within quantum mechanics \cite{om, joos, zur} which is known not to solve the measurement problem. Hence, it is important to distinguish these two types of decoherence by naming them `classical decoherence' and `quantum decoherence' respectively. Quantum decoherence is a purely quantum mechanical process in which a subsytem of a larger system is found in a mixed state when its environment is traced over. Classical decoherence is not a quantum mechanical process. It is driven by the nonlinear term in the classical Schr\"{o}dinger equation which reduces the total system to a mixed state. Such a term is presumably generated when quantum mechanical microsystems aggregate to form classical macrosystems through nanostructural stages. In such processes there is a gradual transition from quantum to classical states brought about by the changing environment of a subsystem. This transition can be phenomenologically described by multiplying the nonlinear term $Q_{cl}$ by a coupling parameter $\lambda$ which varies between 0 and 1 ($0\leq \lambda \leq 1$) as the system makes a transition from the quantum to the classical state \cite{gh1, gh2}. The converse process of isolation of quantum subsystems from their environment is also known to be possible. The intermediate states are {\em hybrid} mesoscopic states that are partly quantum and partly classical, their quantum-ness or classical-ness being parametrized by a value of $0\leq\lambda\leq 1$.

\section{Concluding Remarks}

The operator formulation of classical mechanics presented above goes beyond standard classical statistical mechanics (described by two uncoupled equations) in predicting superpositions of classical states which show no interference pattern: {\em the (complex) sum is more than the parts}. Whether such states really exist is an open question at present. 

We have shown that classical states like (\ref{clsup}) violate macroscopic realism which is widely believed to be an essential characteristic of classical-ness.
This demonstration raises issues concerning certain deeply held notions about {\em realism}, classical-ness and quantum-ness. 

We have also shown that there is no measurement problem if the classical apparatus is described by a classical wave function. 

The method developed above justifies {\em inter alia} the use of the Hilbert space method for classical states in quantum information processing theory. 

Finally, it should be mentioned that complex wave functions for classical systems were first introduced by Koopman \cite{K} and von Neumann \cite{vN} in order to formulate classical statistical mechanics in Hilbert space analogous to quantum mechanics. However, in their formulation the classical wave function $\psi(q,p)$ and its complex conjugate $\psi^*(q,p)$ as well as the phase space density $\rho(q,p) = \psi^*(q,p)\psi(q,p)$ satisfy the classical Liouville equation and all dynamical variables commute. Consequently, the correspondence with quantum mechanics is not as transparent as in the formulation given above.

\section{Acknowledgement}
The basic idea of this paper was presented on 5th December, 2017 at the {\em 3rd Int Conf on Quantum Foundations} organized by NIT, Patna, 4-9 December, 2017. The author is grateful to A. Matzkin for drawing his attention to Refs. (\cite{rosen1}, \cite{rosen2}, \cite{rosen3}) (which predate Refs.\cite{gh1}, \cite{gh2}) and Ref.\cite{ben} after the talk, which necessitated incorporation of some additional material for clarification. The author also acknowledges helpful discussions with Anirban Mukherjee, and thanks the National Academy of Sciences, India for a grant which enabled this work to be undertaken.


\begin{thebibliography}{0}
\bibitem{note}
This crucial aspect of the von Neumann interaction model has not been taken into account in Ref. \cite{rosen3}. Consequently, in \cite{rosen3} the particle (1) and the apparatus (2) remain in a product state after the interaction and are not entangled. In other words, the Hilbert space remains $\mathcal{H}_1 \times \mathcal{H}_2$. Consequently there are no true correlations between the apparatus and the particle.
\end{thebibliography}
\end{document}